
\input phyzzx

\overfullrule=0pt
\font\twelvebf=cmbx12
\nopagenumbers
\footline={\ifnum\pageno>1\hfil\folio\hfil\else\hfil\fi}
\line{\hfil CU-TP-574}
\line{\hfil CERN-TH.6701/92}

\vglue .4in
\centerline {\twelvebf  Vortex Dynamics in Self-Dual Chern-Simons Higgs
Systems}
\vskip .5in
\centerline{\it Yoonbai  Kim$^*$ and Kimyeong Lee$^\dagger$ }

\vskip .1in
\centerline {$^{*\dagger}$Physics Department, Columbia University}
\centerline {New York, New York 10027}
\vskip .2in

\centerline{$^\dagger$Theory Division, CERN}
\centerline{CH-1211 Geneva 23, Switzerland}
\baselineskip=18pt
\overfullrule=0pt

\vskip .5in
\centerline {\bf Abstract}
\vskip .1in
We consider vortex dynamics in self-dual Chern-Simons Higgs systems.
We show that the naive Aharanov-Bohm phase is the inverse of the statistical
phase expected from  the vortex spin, and that the
self-dual configurations  of vortices are degenerate  in energy but not
in angular momentum. We also use the path integral formalism to  derive
the dual formulation of Chern-Simons Higgs
systems in which  vortices appear as charged particles.
We argue that besides the
electromagnetic interaction, there is an additional
interaction between vortices, the so-called Magnus force,
and that these  forces can be put together into  a single
`dual electromagnetic' interaction. This dual electromagnetic
interaction leads  to the right
Aharanov-Bohm phase. We also derive and study
the effective action for slowly moving vortices,
which contains  terms both linear and  quadratic in the  vortex velocity.

\vfill

\noindent October 1992
\footnote{}{$^*$ Y. Kim's address from September  1992  is Physics
Department, Kyung Hee University, Seoul 130-701, Korea  }

\vfill\eject

\def\pr#1#2#3{Phys. Rev. {\bf D#1}, #2 (19#3)}
\def\prl#1#2#3{Phys. Rev. Lett. {\bf #1}, #2 (19#3)}

\def\np#1#2#3{Nucl. Phys. {\bf B#1}, #2 (19#3)}
\def\pl#1#2#3{Phys. Lett. {\bf #1B}, #2 (19#3)}

\REF\rSELF{ J. Hong, Y. Kim, and P.Y. Pac, \prl{64}{2330}{90};
R. Jackiw and E.J. Weinberg, \prl{64}{2334}{90}.}

\REF\rJLW{ R. Jackiw, K. Lee, and E. Weinberg, \pr{42}{3488}{90}.}

\REF\rANY{F. Wilczek  \prl{48}{1144}{82}; F. Wilczek and A. Zee,
\prl{51}{2250}{83}.}

\REF\rVORT{J. Fr\"ohlich and P.A. Marchetti, Comm. Math. Phys. {\bf 121},
177 (1989).}

\REF\rAnyon{ A.S. Goldhaber and R. Mackenzie, \pr{214}{471}{88}; T.H.
Hansson, M. Ro\v{c}ek, I. Zaed and S.C. Zang, \pr{214}{475}{88};
A.S. Goldhaber, R. Mackenzie and F. Wilczek, Mod. Phys. Lett. {\bf A4},
21 (1989).}

\REF\rDUAL{ P.K. Townsend and P. van Nieuwenhuizen,
\pl{136}{38}{84}; S. Deser and  R. Jackiw, \pl{139}{371}{84};
 X.G. Wen and A. Zee, \prl{62}{1937}{89}; T.H. Hansson and A. Karlhede,
Mod. Phys. Lett. {\bf A4}, 1937 (1989).}

\REF\rMagnus{ R.L. Davis and E.P.S. Shellard, \prl{63}{2021}{89};
 R.L. Davis,  \pr{40}{4033}{89}; R.L. Davis,
Mod. Phys. Lett. {\bf A5}, 853 (1990).}

\REF\rFluid{ E.B. Sonin, Rev. Mod. Phys. {\bf 59}, 87 (1987). }

\REF\rManton{ N.S. Manton, \pl{110}{54}{82}.}

\REF\rVORT{  R. Ruback. \np{296}{669}{88};
K.J.M. Moriaty, E. Meyers and C. Rebbi, \pl{207}{411}{88}; E.P.S.
Shellard and P.J. Ruback, \pl{209}{262}{88};
T.M. Samols, Comm. Math. Phys. {\bf 145}, 149 (1992); E. Myers, C. Rebbi
and R. Strika, \pr{45}{1355}{92}.}

\REF\rWang{R. Wang, Comm. Math. Phys. {\bf 137}, 587 (1991).}

\REF\rLATT{A. Savit, Rev. Mod. Phys. {\bf 52}, 453 (1980).}

\REF\rMCSH{ S. J. Rey and A. Zee, \np{352}{897}{91}.}

\REF\rSta{S.K. Kim and H.-S. Min, \pl{281}{81}{92}.}

\REF\rLeese{R.A. Leese, \np{366}{283}{91}.}

\chapter{Introduction}

Recently, several  studies of
self-dual abelian Chern-Simons Higgs
systems in 2+1 dimensions have appeared.\refmark{1,2} These
 self-dual models have a
specific sixth-order potential which has degenerate symmetric and
asymmetric vacua. There is a Bogomol'nyi-type bound on the energy
functional, which is  saturated by  configurations satisfying certain
first-order equations.  These self-dual configurations consist of
 topologically stable vortices in the asymmetric phase, and
nontopological solitons in the symmetric phase.  These solitons carry
both electric charge and magnetic flux, resulting in nontrivial
spin, and can be regarded as anyons, or
particles with  fractional spin and statistics.\refmark{3}

While attention has been paid to  the statistics of
vortices in the asymmetric phase,\refmark{4,5} there are
many aspects of vortices which still need to be understood clearly.
One aspect is  the spin-statistics for vortices. Another
is the dynamics of slowly moving vortices in  self-dual
systems.  In this paper, we study  various questions related to
vortices in Chern-Simons Higgs
systems.

Let us start first with considering the angular momentum of
nontopological solitons and vortices. One striking fact is that for a
given charge or magnetic flux, the angular momentum of nontopological
solitons without any vorticity in the symmetrics phase
 has the opposite sign compared with  that of topological
vortices in the asymmetric phase.\refmark{2}  The
 spin-statistics theorem implies that the
spin of a particle is directly related to the statistics of
that particle.
The statistics of  elementary charged particles\refmark{5} and nontopological
solitons are determined  by the Aharanov-Bohm phase
due to electric charge and magnetic flux. Since vortices could
have the same charge and magnetic flux but opposite spin compare with
nontopological solitons, the statistics of vortices cannot be explained
by the naive Aharanov-Bohm phase.  This is the first puzzle we will
consider.

A self-dual configuration of $n$ vortices appears   to be completely
specified  by the vortex  positions, that is, by
 $2n$ real parameters.\refmark{1,2}
As we will see, all configurations of given number of vortices are
 degenerate  in energy but not in angular
momentum.  For a  system of
two vortices, total angular momentum decreases from four times the
vortex spin to twice the vortex spin as their separation
increases
from zero to infinity. The influence of this change on the motion of
vortices with small kinetic energy is the second puzzle we shall
consider.

In order to understand the statistics of vortices, we reformulate the original
theory in a way that makes the interaction between vortices manifest.
This reformulation is called the dual formulation, where the
massive vector boson in the asymmetric phase is described by the
Maxwell and Chern-Simons  rather than Chern-Simons and Higgs
terms, and where  vortices appear as charged particles. The dual formulation
has been derived many times in past using the
equations of motion or a lattice model.\refmark{6} We present here a
clearer  derivation using the path integral formalism.

Some physical implications  of the dual formulation of various
three dimensional field theories have  been studied previously.\refmark{7}
In the theory of a complex scalar field with a
global abelian symmetry, a vortex in a uniform charged background
feels the so-called Magnus force, which has more-or-less the same
origin as the force responsible for the curved flight of a spinning ball.
The Magnus force on a curve ball is proportional to its speed
and is  perpendicular to its direction, very much like the Lorentz force.
In the dual formulation, vortices are charged particles and the background
charge density becomes a magnetic field. The Magnus force on vortices
 becomes a Lorentz force.
The concept of the Magnus force is important for vortex
dynamics in superfluids.\refmark{8}
One can also see the Magnus force for
vortices in Maxwell Higgs theories  when there is a
   background electric charge density
 screened by the Higgs field.  While it is  possible to see
the Magnus force in the original formulation, it appears more
transparently in the dual formulation.

In Chern-Simons Higgs systems,  vortices  carry both magnetic
flux and electric charge.
  Because vortices feel the charge of other vortices,
  vortices also feel also the Magnus force in the absence of a
 background charge.
In the dual formulation  both the
electromagnetic and Magnus forces are combined  into a single
`dual electromagnetic' force. The Aharanov-Bohm phase in the dual formalism
determines  the statistics of vortices and yields
 exactly what one expects given  the vortex spin.

The total angular momentum of many overlapping
vortices  is
equal to the vortex spin times the square of the total vorticity. When
vortices at rest are separated from each other by a  distance much larger
than the vortex core size, one finds that the total angular momentum
is  just the sum of the  individual vortex spins. The physical reason is that
any gauge-invariant local field falls off exponentially to its  vacuum
configuration as one moves away from any vortex. For  self-dual
vortex configurations  characterized by the positions of the
vortices, the total energy is just the sum of the individual vortex masses
but the total angular momentum is a  function of the vortex positions.

The behavior of the total angular momentum could be understood as
follows.   Consider two noninteracting point  anyons of spin $s$ in
nonrelativistic quantum mechanics. Two
separated anyons at rest have zero classical orbital angular momentum.
Quantum mechanically, orbital angular momentum is given by
$2s + 2\hbar \times {\rm integer}$, which in turn implies that the
total angular momentum is $4s+ 2\hbar\times {\rm integer}$.  Our
vortices are extended objects and can overlap each other.
If we quantize the self-dual configurations of two vortices,
there would be many states whose orbital angular momentum
varies from $2s$ to $ 2s+ 2\hbar \times {\rm integer}
\sim 0$ with the same energy.
The average separation of two vortices in these states will
increase as the orbital angular momentum decreases.

We are also interested in how the position dependence of  the total
angular momentum affects the classical dynamics  of slowly
moving vortices in self-dual Chern-Simons Higgs systems. To understand
the dynamics of vortices in general,
we derive  the effective lagrangian for slowly moving
vortices. We follow  Manton's approach\refmark{9} which means for ur
case  that for a given number of vortices,
the field configurations of slowly moving vortices are  very
similar to the field configurations of vortices at rest and the
effective action for slowly moving vortices is determined by the
characteristics of the self-dual configurations of vortices.

We approach the problem from the lagrangian point of view. We imagine that the
motion of slowly moving vortices is a generalization of the
nonrelativistic limit of the Lorentz transformation.
This implies that  the field
 configurations of vortices in motion satisfy   the field equations to
first order in the vortex velocity. We  calculate the
field-theoretic lagrangian, which then yields the effective action  for
those vortices  as a functional of the vortex positions and velocities.
We show that the orbital angular
momentum for vortices at rest calculated from the
 effective action is
identical to that calculated from the field theory.

The contents of this paper are as follow. In Sec.~2, we briefly
 review
vortex configurations in self-dual Chern-Simons Higgs systems. We
analyze  the total angular momentum of vortices at rest and present
the result of a numerical analysis for two vortices at  finite
separation.  In Sec.~3 we present the dual transformation of
Chern-Simons Higgs theories  in the path-integral formalism. Here
we include external currents and fields in the transformation.  In
Sec.~4, we study various aspects of the dual formulation. We relate the
statistics of vortices to  the Magnus force. We also discuss the effect
of external currents and fields in the dual formulation.
In Sec.~5, we derive  and study the effective  lagrangian of slowly
moving  vortices.
 In Sec.~6, we
conclude with some remarks.  In appendix A, we present the dual
formulation  of the theory of a complex scalar field with a global
abelian symmetry and discuss the Magnus force. In appendix B, we
present the dual formulation of Maxwell Higgs theories.
 In appendix C, we   derive the effective lagrangian for  slowly moving
vortices of   self-dual Maxwell Higgs systems using
 the dual formulation of appendix B. This effective lagrangian has been
studied in detail both numerically and analytically by various
authors.\refmark{10}
\endpage

\chapter{Model}

We consider the theory of a complex scalar field
 $\phi = fe^{i\theta} /\sqrt{2}$  interacting with a
gauge field $A_\mu$ whose kinetic term is the Chern-Simons term. The lagrangian
for the theory is given by
$$ {\cal L} = {\kappa \over 2}\epsilon^{\mu\nu\rho}A_\mu\partial_\nu
A_\rho +
{ 1\over 2}
(\partial_\mu f)^2 + {1\over 2} f^2 (\partial_\mu \theta +
A_\mu ) ^2 - U(f).  \eqno\eq $$
Self-dual models arise when the potential has the form,
$$ U(f)  = {1\over  8\kappa^2} f^2  ( f^2 - v^2)^2. \eqno\eq $$
Gauss's law constraint obtained from the variation of $A_0$ is given by
$$ \kappa F_{12} + f^2 (\dot{\theta} + A_0) = 0,
\eqno\eq $$
where dot denotes the time derivative.  Gauss's  law implies that
the total magnetic flux $\Psi = \int d^2r F_{12}$ and
the total electric charge $Q =  \int d^2r  f^2 (\dot{\theta} + A_0) $
are related by
$$\kappa \Psi = - Q. \eqno\eq $$

For  the self-dual models  the energy functional can be written as
$$ \eqalign{ E = &\ \int d^2r \biggl\{ {1\over 2} \dot{f}^2 + {1\over 2}
[ \partial_i f \mp \epsilon_{ij}
f(\partial_j \theta + A_j) ]^2 + {1\over 2}
f^2[ (\dot{\theta} + A_0) \pm {1\over 2\kappa}
(f^2 - v^2)]^2 \biggr\} \cr
&\  \pm m_p Q,\cr}
 \eqno\eq $$
where $ m_p = v^2 /2\kappa$ is the mass of charged particles in the symmetric
phase. As the integral in Eq.(2.5) is positive, there is  a bound on
the nergy functional,
$$ E \ge m_p|Q|. \eqno\eq $$
This bound is saturated by the configurations satisfying
$$ \eqalign{  &\ \dot{f} = 0, \cr
&\  \partial_i f \mp \epsilon_{ij} f(\partial_j \theta + A_j)
= 0  , \cr
 &\ \dot{\theta} + A_0 \pm  {1\over 2\kappa }   (f^2 - v^2 ) = 0,
 \cr }
\eqno\eq $$
and Gauss's  law (2.3). In the remainder of this section,
  we will consider only the positively
charged configurations.

If there are vortices of unit vorticity at points  $\vec{q}_a$, $
 a= 1,,,n$,  the phase variable can be chosen to be
$$ \theta = \sum_{a=1}^n {\rm Arg} (\vec{r}-\vec{q}_a), \eqno\eq $$
which satisfies  $\epsilon_{ij}\partial_i \partial_j
\theta = 2\pi \sum_a \delta(\vec{r}-\vec{q}_a)$. Eq.(2.7) implies that
the total magnetic flux is given by $\Psi = -2\pi n$ for this
configuration.
Eqs.(2.3), (2.7) and (2.8) imply that the $f$ field satisfies
$$ \partial_i^2 \ln f^2 - {1\over \kappa^2}f^2(f^2-v^2) = 4\pi \sum_a
\delta(\vec{r}-\vec{q}_a).
\eqno\eq $$
Assuming that vortices  are not overlapping, we can analyze the behavior
of the $f$ field near  $\vec{q}_a$. In the complex
coordinate of positions, Eq.(2.9) implies
$$ \eqalign{ \ln f^2 = &\ \ln |z-q_a|^2 + c+   b_1(z-q_a) + b_2(z-q_a)^2
+b_3(z-q_a)^3 \cr
&\ +b_4(z-q_a)^4 + b^*_1(z^*-q_a^*)+b^*_2(z^*-q^*_a)^2 +
b^*_3(z^*-q^*_a)^3\cr
&\ + b^*_4(z^*-q^*_a)^4 - {v^2\over 16 \kappa^2}e^c |z-q_a|^4
+ O((z-q_a)^5),\cr} \eqno\eq $$
where real $c[\vec{q}_a]$ and complex $b_i[\vec{q}_a]$   are
defined with respect to $\vec{q}_a$ and functions of the positions of
other vortices.

In addition to magnetic flux  and energy, there is angular momentum
which characterizes  a given configuration.
The angular momentum functional  $J = \int d^2x \epsilon_{ij} r^i T^{0j}$
is given by
$$ \eqalign{ J &\ =  - \int d^2r \epsilon_{ij}r^i \bigl[ \dot{f}
\partial_j f + f^2 (\dot{\theta}
+ A_0)(\partial_j \theta + A_j) \bigr] \cr
&\ = - \int d^2r \epsilon_{ij}r^i\bigl[ \dot{f}\partial_jf - \kappa F_{12}
(\partial_j \theta + A_j) \bigr] \cr} \eqno\eq $$
with Gauss's law (2.3).

Any rotationally symmetric solution with vorticity $n$
in the cylindrical  coordinate $(r,\varphi)$
would be given by the ansatz, $f(r), \theta = n\varphi $, and  $
A_i = \epsilon_{ij} r^j (a(r) - n ) /r^2 $. There are two finite  energy
solutions of Eq.(2.7) with this ansatz. When $f(\infty) = 0 $, the
solution becomes a nontopological soliton with vorticity $n$.
 In this case
we can have $a(\infty) = -\alpha$ where $\alpha > n+ 2$.
When $f(\infty) = v$,  $a(\infty) = 0$ and  the solution becomes
overlapped vortices. The total magnetic flux  of  this ansatz is
$$ \Psi = -2\pi (n + \alpha). \eqno\eq$$
The angular momentum of the solution can be calculated from
Eq.(2.11),\refmark{2} leading to
$$ J = \pi \kappa (\alpha^2 - n^2).\eqno\eq $$
Since we have used just Gauss's law,  Eq.(2.13) is applicable to
theories with  more general potential than the self-dual one (2.2).
Nontopological solitons in the symmetric phase have the energy per charge
identical to that of elementary charged particles, implying that they are
at  the verge  of instability. Vortices in the asymmetric phase are however
stable for the topological reasons.

In the symmetric phase elementary
 particles have spin $s_p = 1/4\pi \kappa$
and nontopological solitons has spin $s_p Q^2 - Qn$. Since total
charge would be quantized in integers , the $Qn$ part would be an integer.
The statistics of elementary charged particles and nontopological
solitons is given by the phase change  of the wave function for two identical
objects when they are rotated by 180 degrees counter-clockwise wise around
the center of mass, which will be $e^{2\pi i s_p Q^2}$ and is identical
to the half of the  Aharanov-Bohm phase,
$ e^{i\Psi Q/2}$. On the other hand, vortices of unit
voticity would carry spin $s_v = -\pi\kappa$ and the correct
statistics would be $e^{2\pi i s_v} = e^{-2\pi i s_pQ^2} $, which
cannot be the naive Aharanov-Bohm phase.

The self-dual configurations of vortices seem
to be  parameterized only by the positions
$\vec{q}_a$. Energy is independent of vortex positions and so the
derivative of fields with respect to vortex positions would become
$2n$ zero-modes of the self-dual equations. For  self-dual
configurations, the total angular momentum (2.11) becomes
$$ J = \kappa \int d^2r \epsilon_{ij}r^i (\partial_j \theta + A_j)
F_{12}.\eqno\eq $$
{}From Eq.(2.3),
we know that $F_{12}$ vanishes at points $\vec{q}_a$'s because $f^2$ vanishes
there. Without changing the value of $J$, we can  then subtract these
points from the integration in Eq.(2.14). Except at these points, $
F_{12} = \partial_1
\bar{A}_2 - \partial_2 \bar{A}_1$ with $\bar{A}_i = A_i + \partial_i
\theta$. Noting that
$$ \epsilon_{ij}r^i \bar{A}_j \epsilon_{kl}\partial_k \bar{A}_l
= \partial_i \biggl[ {1\over 2}r^i (\bar{A}_j)^2 - \bar{A}_i
(r^j\bar{A}_j) \biggr] + (\partial_i\bar{A}_i)(r^j\bar{A}_j),\eqno\eq $$
and $\bar{A}_i$ being  transverse from Eq.(2.7),
we can write the angular momentum as a boundary integration,\refmark{2}
$$ J = - \kappa \biggr[ \sum_a \oint dl^i_a  - \oint_\infty dl^i
\biggr] \epsilon_{ij} \left\{
{1\over 2} r^j (\bar{A}_k)^2 - \bar{A}_j (r^k \bar{A}_k) \right\},
\eqno\eq
$$
where the sum is over the positions of vortices and the line intergal
is around a small counter-clockwise circle
 around $\vec{q}_a$. There is no spatial
infinity term in the asymmetric phase. (For nontopological solitons
 the boundary at spatial infinity contributes
and this contribution does not depend on the position or shape of
solitons.)

Let us evaluate the integral in Eq.(2.16)  at each  vortex
position
$\vec{q}_a$. Near $\vec{q}_a$, we can put $r^i  = q_a^i +
\epsilon_{ij}  l_a^j $ and Eq.(2.16)  becomes
$$ \eqalign{ J =&\  -\kappa \sum \oint dl^i_a \epsilon_{ij}\left\{
{1\over 2} q^j_a (\bar{A}_k)^2 - \bar{A}_j (q^k_a \bar{A}_k)\right\}
\cr
&\ -\kappa \sum \oint dl^i_a \epsilon_{ij}\left\{
{1\over 2}\epsilon_{jk} l^k_a (\bar{A}_l)^2 - \bar{A}_j
(\epsilon_{kl}l^l_a \bar{A}_k)\right\}.\cr}\eqno\eq $$
We use  Eqs.(2.7) and (2.10) to expand $\bar{A}_i$ near $\vec{q}_a$,
$$ \eqalign{ \bar{A}_i &\ = -\epsilon_{ij}\partial_j \ln f \cr
 &\ = - \epsilon_{ij}\left\{ \epsilon_{jk}{l^k_a \over |\vec{l}_a|^2} +
b^j_1 + O(l_a) \right\},\cr } \eqno\eq $$
where $b_1 = b_1^1 -i b_1^2$. We evaluate the integration
in Eq.(2.17)  with  Eq.(2.18) to get the total angular momentum as
$$ J = - 2\pi \kappa \sum_a \vec{q}_a \cdot \vec{b}_1[\vec{q}_a] - \pi
\kappa |n| \eqno\eq $$
with  total voticity $n$. The first term of the right hand side of
Eq.(2.19) represents the orbital part and the second
term represents the spin part.

There is  only one length scale, $v^2/\kappa$, in the problem.
As the distance  between vortices goes
 to infinity, the field around each
vortex position would approach to the rotationally symmetric ansatz
exponentially, which with Eq.(2.10) means that $b_i[q_a]$'s vanish
exponentially with the
 mutual distance. Thus the total angular momentum (2.19)
would changes from $s_v n^2$ to $s_v n$ as  vortices get separated
from each other. The self-dual vortex configurations are
degenerate in energy but not in angular momentum.

\FIG\fone{ Plot of the $f$  field in unit of $v$ on the
$x-y$ plane for two vortices of mutual distance $d=6$  with spatial
distance unit $v^2/\kappa$ }
\FIG\ftwo{ Plot of the magnetic field  $F_{12}$ in  unit of
$v^4/4\kappa^2$ on the $x-y$ plane with $d=6$ }
\FIG\fthree{ Plot of the total angular momentum in unit of
$-\pi\kappa$ as a function of mutual distance $d$.}

Let us investigate briefly the self-dual configurations of two vortices
by a numerical analysis.  Although the existence of multi-vortex
solutions is proved,\refmark{11} no exact analytic solutions is found.
 Fig. 1 and 2  show the magnitude of the scalar field and the
magnetic field  at $d = 6 \kappa /v^2 $.
In Fig. 3, we show  the total angular momentum as a function of
the separation distance. Angular momentum decreases from $-4\pi
\kappa$ to $-2\pi\kappa$, supporting the argument in the previous
paragraphs.

\endpage

\chapter{Dual Formulation}

Suppose we are interested in the calculating physical amplitude
by using the path integral
formalism.  For a generality we include   an external gauge field
$A^{ext}_\mu$ and  current $J^\mu$. The lagrangian is then
$$ {\cal L} = {\kappa \over 2} \epsilon^{\mu\nu\rho} A_\mu \partial_\nu A_\rho
+ {1\over 2} (\partial_\mu f)^2 + {1\over 2} f^2 (\partial_\mu \theta +
 A_\mu + A^{ext}_\mu)^2 - U(f)  + A_\mu J^\mu .
 \eqno\eq $$
The generating functional will be
$$ Z= <F| e^{-iHT}|I> = \int [df][d\theta][dA_\mu]\prod_x f(x)
\bar{\Psi}_F \exp \{ i \int d^3x {\cal L} \} \Psi_I,\eqno\eq $$
where there is a  nontrivial Jacobian factor because we use the radial
coordinate for  the scalar field. The initial and final wave functions,
$\Psi_{F,I}$ give the  necessary boundary conditions.

A given field configuration in the path integral could contain vortices
and antivortices and  the $\theta$ field could be   multivalued.
We can in principle split the  $\theta$  field into two parts,
$$ \theta(\vec{r}, t)  = \bar{\theta}(\vec{r},t)
 + \eta(\vec{x},t),\eqno\eq $$
where the first term describes a configuration of vortices,
$$ \bar{\theta}(\vec{r},t) = \sum_a (-1)^a {\rm Arg} ( \vec{r} - \vec{q}_a(t))
\eqno\eq$$
with  vorticities   $(-1)^a$ and locations $\vec{q}_a(t)$, and
the second term $\eta$ represents  single-valued fluctuations around a given
configuration of vortices. From the multi-valued  $\bar{\theta}$,
we can construct the vortex current,
$$ \eqalign{ K^\mu(x) &\ \equiv  {1 \over 2\pi} \epsilon^{\mu\nu\rho}
\partial_\nu  \partial_\rho \bar{\theta}  \cr
&\ = \sum_a (-1)^a
( 1, {d\vec{q}_a \over dt} )\delta^2(\vec{r} - \vec{q}_a(t)) \cr
&\ = \sum_a  (-1)^a \int d\tau {dq^\mu_a \over d\tau}
\delta^3 (x^\nu  - q_a^\nu(\tau)),\cr}
\eqno\eq $$
which satisfies the conservation law, $\partial_\mu K^\mu = 0$.
Integration over the $\theta$  variable becomes
$$  [d\theta] = [d\bar{\theta}][d\eta]= [dq^\mu_a] [d\eta],\eqno\eq  $$
which means that we sum    over single-valued fluctuations around a given
configuration of vortices and then
sum over all possible configurations of
vortices,  including annihilation and creation of vortex pairs.

Let us now linearize the third term of the lagrangian (3.1) by introducing an
auxiliary vector field $C^\mu$,
$$ \eqalign{ &\  \prod_x f(x) \exp  i\int d^3x \bigl[ {1\over 2}f^2
(\partial_\mu \theta + A_\mu + A_\mu^{ext})^2 \bigr] \cr
&\ =
 \int [dC^\mu] \exp  i\int d^3x \biggl\{ -{1
\over 2 f^2} (C^\mu)^2 + C^\mu (\partial_\mu \bar{\theta} +
\partial_\mu \eta +A_\mu + A^{ext}_\mu )
 \biggr\},
\cr} \eqno\eq $$
where the nontrivial Jacobian is essential.
As $\eta$ is single-valued, one can integrate over
  $\eta$ in the standard way, leading to
$$ \int [d\eta] \exp \biggr[ i\int d^3x C^\mu \partial_\mu \eta \biggl]
= \delta(\partial_\mu C^\mu).\eqno\eq $$

Now we introduce the dual gauge field $H_\mu$ to satisfy
$$ \int [dC^\mu]  \delta(\partial_\mu C^\mu)...
 =\int [dC^\mu][H_\mu]    \delta(
  C^\mu - {1\over 2\pi}
\epsilon^{\mu\nu\rho} \partial_\nu H_\rho  )...
\eqno\eq $$
where the dots denote the integrand.  There would be infinite
gauge volume which can be taken care of later, but there is
no nontrivial
Jacobian factor as the change of variables  is linear.
By  using the fact
$$ {1\over 2\pi} \epsilon^{\mu\nu\rho} (\partial_\mu \bar{\theta})\partial_\nu
H
  = K^\mu H_\mu \eqno\eq$$
up to a total derivative, we  can integrate
over $C^\mu$, resulting in the lagrangian,
which is
$$
\eqalign{ {\cal L}' = &\
-{1\over 16\pi^2 f^2} H_{\mu\nu}^2 + H_\mu K^\mu +{1\over 2}
 \epsilon^{\mu\nu\rho} A_\mu \partial_\nu ( \kappa A_\rho +
{1\over \pi}H_\rho) \cr
&\ +{1\over 2\pi}\epsilon^{\mu\nu\rho} H_\mu \partial_\nu
A^{ext}_\rho + A_\mu J^\mu +...\cr} \eqno\eq $$
where $H_{\mu\nu} = \partial_\mu H_\nu - \partial_\nu H_\mu $ and
the dots indicate $f$-dependent terms.

The exponent is quadratic in $A_\mu$ and so the integral over
$A_\mu $ is easy.  The equation of motion for $A_\mu$ is
$$  \kappa \epsilon^{\mu\nu\rho} \partial_\nu A _\rho = - {1\over 2\pi}
\epsilon^{\mu\nu\rho} \partial_\nu H_\rho - J^\mu,
\eqno\eq $$
whose solution is formally
$$ A_\mu =- {1\over 2\pi \kappa} H_\mu + V_\mu,\eqno\eq $$
where $\epsilon^{\mu\nu\rho}\partial_\nu V_\rho = - J^\mu /\kappa$.
Rather than integrating   over $A_\mu$,
we  substitute $A_\mu$ in the path integral by Eq. (3.13) and then
the integration over  $A_\mu$ becomes the integration over $V_\mu$.

The resulting path integral becomes
$$ Z = \int [df][dq^\mu_a]
[dH_\mu][dV_\mu] \exp i\int d^3x {\cal L}_{D},\eqno\eq
$$
where the dual transformed lagrangian is
$$ \eqalign{ {\cal L}_{D} =  &\ {1\over 2} (\partial_\mu f)^2 - U(f)
-{1\over 16\pi^2 f^2} H_{\mu\nu}^2 -{1\over 8\pi^2 \kappa}
\epsilon^{\mu\nu\rho}H_\mu \partial_\nu H_\rho  + H_\mu K^\mu \cr
&\ -  {1\over 2\pi \kappa} H_\mu J^\mu  +{1\over 2\pi}
\epsilon^{\mu\nu\rho}H_\mu \partial_\nu A^{ext}_\rho
 + {\kappa \over 2} \epsilon^{\mu\nu\rho}V_\mu \partial_\nu V_\rho +
V_\mu J^\mu. \cr }
\eqno\eq $$
There is no Jacobian factor in the measure. One can introduce the
gauge fixing terms for $H_\mu$ and $V_\mu$.
 The sign difference
between the Chern-Simons terms of the original and dual transformed
theories will be crucial in understanding the statistics of vortices.
 The original gauge field is separated into two pieces, $H_\mu$ and
$V_\mu$. The vortex current $K^\mu$ becomes an electric current for the
dual field $H_\mu$. External current is however coupled to both
the dual gauge field $H_\mu$ and reduced gauge field $V_\mu$.

The mass  of vortices arises from the cloud of the $f,H_\mu$ fields
surrounding them     is finite.
The variation of $H_0$ implies the Gauss's law constraint,
$$ \partial_i({1\over f^2}H_{0i}) - {1\over \kappa}H_{12} + 4\pi^2 K^0 +
2\pi F_{12}^{ext} - {2\pi \over \kappa}J^0 = 0,\eqno\eq $$
which would dictate the cloud around vortices.

The relation between the original fields and dual fields in the classical
level  can be seen from the   equations one would get from the
lagrangians at each step.    They are related to
each other by
$$ f^2(\partial_\mu \theta + A_\mu + A_\mu^{ext}) = C_\mu =
{1\over 2\pi} \epsilon_{\mu\nu\rho}
\partial^\nu H^\rho, \eqno\eq $$
Another relation between the original and dual fields is given by
Eq.(3.13). The original $U(1)$ charge  is then given by
$$ \eqalign{ Q &\ = \int d^2r \left\{
f^2(\dot{\theta}  + A_0 + A_0^{ext} )  + J^0 \right\} \cr
&\  = \int d^2r \left\{ {1\over 2\pi}H_{12} + J^0 \right\}  \cr
&\ =  \int d^2r \left\{  2\pi\kappa K^0
+ {\kappa \over 2\pi } \partial_i ( {1\over f^2} H_{0i})
+ \kappa F^{ext}_{12}  \right\},\cr} \eqno\eq $$
where the last equality comes from Gauss's law (3.16).
The second-to-last term gives nonzero contribution to the charge for
the configuration of nontopological solitons in the symmetric phase
where
$f$ field vanishes only on isolated points and the spatial infinity.
Note that the charge conservation in dual formulation is satisfied
by the topology of the field configuration, not by the field equations.

\endpage

\chapter{Physical Consequences}

We have obtained the dual formulation of Chern-Simons Higgs
systems,  which could  be useful in
understanding the various physical aspects of the asymmetric phase.
 In the dual formalism the interaction between
vortices is more direct because they appear as charged particles rather
than topological objects. In addition, we can see the
interaction between
vortices and external currents and fields more directly.
The dual transformation in general transforms the
weak coupling  into the strong coupling
 and vice versa and
 has been widely used to understand the phase structure of
a given theory. (See Ref.[12] for a review.)
  If we try to quantize vortices by the semiclassical method,
  the coupling between elementary particles should be
very small, or $\kappa \gg 1$ for the method to be a good
approximation.   In this case, vortices interact with each other
strongly as one can see from the dual formulation.
However this aspect of the dual formulation would not be explored
in this paper. Let us now make a few
observation  about the dual formulation.

\noindent 1) {\it Massive vector bosons}

There are two ways to describe a massive vector boson of spin one in three
dimensions: the  Maxwell and Chern-Simons terms, or the Chern-Simons
and Higgs terms. This observation led       to the original derivation of
the dual transformation.\refmark{6}  From Eqs.(3.1) and (3.15) with $f=v$
we have two equivalent lagrangians,
$$ \eqalign{ &\
{\cal L}_1 = {\kappa \over 2}\epsilon^{\mu\nu\rho} A_\mu
 \partial_\nu A_\rho  + {1\over 2} v^2 A_\mu^2, \cr
&\    {\cal L}_2 = -{1\over 16\pi^2 v^2} H_{\mu\nu}^2
- {1\over 8\pi^2 \kappa} \epsilon^{\mu\nu\rho}
H_\mu \partial_\nu H_\rho. \cr} \eqno\eq $$
Both describe a particle of mass $m = v^2/|\kappa|$
and spin $-\kappa/|\kappa|$

\noindent 2) {\it Quantum Magnus Phase}

We know how a spinning baseball curves. Let us consider a two-dimensional
version.  When a ball is moving to the negative
$x$ direction with clockwise rotation
in a fluid, the wind velocity (at the ball's rest frame)
on the positive $y$ part is
faster than that on the negative $y$ part, resulting in the pressure
difference. The net force on the ball is then pointing the positive $y$
direction. The magnitude of the force is proportional to the ball velocity
and so this force is somewhat similar to an effective Lorentz force
due to a constant magnetic field. When the moving object is a vortex, it is
called the Magnus force.\refmark{8}

The simplest example in field theory   is
the theory of a complex scalar field with a global abelian symmetry.
When global vortices in this theory move in a unform charge background,
it feels this Magnus force. As one can see in appendix A, the
charge density appears as a uniform magnetic field and vortices as
charged particles in the dual formulation. The Magnus force is given
exactly as a Lorentz force.

When  vortices in the asymmetric phase of
a  Maxwell Higgs theory interact  each other
without any background charge, we know that there is no Magnus force.
As shown in appendix B, vortices again become charged particles in the
dual formulation.  The  background charge    again
appears as a uniform external
magnetic field in the dual formulation
     and vortices in such background would feel
the Magnus force.

Vortices in a   Chern-Simons Higgs theory carry both magnetic flux and
charge around their core. When vortices are close to each other they
would feel the electromagnetic force, which leads to an Aharanov-Bohm
phase in large distance. Because vortices carry charge, there would be
the Magnus force between vortices. The Magnus force is like a Lorentz
force in nature and would lead to an additional Aharanov-Bohm phase.
The total Aharanov-Bohm phase would then be a sum of those from
these two forces.
As we have seen in the previous section, the dual
formulation of Chern-Simons Higgs theories has  the
dual gauge interaction
between vortices. Thus, one would say the original electromagnetic force
and the Magnus force come together as a single dual gauge force.
Vortices
are charged particles in the dual formulation and so their statistics
should be given by the Aharanov-Bohm phase coming from the dual
lagrangian (3.15).
The Aharanov-Bohm phase is a long distance effect
and determined only by the Chern-Simons term.
The sign of the Chern-Simons term in the dual
lagrangian is different from that of the original lagrangian,
which makes the real Aharanov-Bohm phase between vortices to be exactly
inverse of the naive Aharanov-Bohm phase in the original formulation.
This implies that the Magnus force is two times larger than and has
the opposite sign to the original electromagnetic force.
The statistics from the dual Aharanov-Bohm phase is consistent with
what we expect from the vortex spin.
Although the dual formulation has been discovered  many
times,\refmark{6}  this aspect of vortex interaction has not been
noticed.

\noindent 3) {\it External Current and Field }

We have derived the dual lagrangian which is valid even with
external currents  and fields.  Let us see
first  the effect of the external
point charge. In the asymmetric phase
of a Maxwell Higgs theory, any charge will be completely screened and
the net total charge is zero. In the asymmetric phase of a
Chern-Simons Higgs theory, the charge of vortices cannot however
be screened  because  Gauss's law (2.3) or (3.16) implies that
total charge is nonzero when there is nonzero magnetic flux.
If there is no external field and vortex, there cannot be any net
magnetic flux for any finite energy configurations and so total
charge is zero implying that external charges are totally screened.
For a given external point charge, the screening charge will surround
 this charge with the length scale given from the Higgs mass.

What is the interaction between external currents  and vortices? The
dual lagrangian (3.15) leads to the answer. Both of them are charged
currents of the dual gauge field and so there would be nontrivial phase
when the external charge goes around a vortex in a full circle.
This phase is determined by the dual lagrangian and  is given by
$e^{-i2\pi Q_{ext}}$ with external charge $Q_{ext}$.
If the charge is fractional, the phase is nontrivial
 because the Higgs field carries a unit charge
and can screen only integer charges completely.
{}From Gauss's law (3.16) we see that a uniform external charge density
is screened by  a uniform dual magnetic field.
A single vortex moving on this background would feel
the Magnus force which appears as a Lorentz force.

What is the interaction between two external point charges? They are
interacting through two gauge fields $H_\mu$ and $V_\mu$. In the large
distance, the Aharanov-Bohm phases due to two gauge interaction would
 cancel each other,  and there is no nontrivial phase between them.
 Since the screening charge has a finite core size, in a short distance
 external charges would see the nontrivial statistics.

We can ask whether the external currents  and fields could have a
dynamical origin. In Sec. 3  we have not used the
 conservation of external currents explicitly. The key aspect is
that  external currents
 and the gauge field $A_\mu$ couple linearly.
 There seems to be two simple  examples where external currents arise
dynamically.  The matter  could be made of
fermions, in which case $A_\mu J^\mu $ would  be replaced by
$$ {\cal L}_F =  i \bar{\psi}\gamma^\mu (\partial_\mu +ieA_\mu)\psi +
...                \eqno\eq $$
where the dots denote the mass and Yukawa interaction terms.
Or, the matter field could be made of a simple gauge field, in which
case the additional lagrangian would be
$$ {\cal L}_W = e
\epsilon^{\mu\nu\rho} A_\mu \partial_\nu  W_\rho
+...                      \eqno\eq $$
where the dots indicate the kinetic terms for the $W_\mu$ field.
The external field $A^{ext}_\mu$ can be made dynamical by replacing
$A^{ext}_\mu$ in Eq. (3.1) by
a gauge field  $W_\mu$  with some kinetic
term. It is trivial to see how these dynamical degrees couple to the
dual gauge field, which we will not bother to write down.

One may wonder whether there is any dual formulation of Maxwell Chern-Simons
Higgs theories. One can follow the similar procedure as in Sec.3 and appendix B
 and will end with a dual formulation with two gauge fields
  even when there is
no external currents and fields. In the case where the $f$ field is
a constant, a dual formulation with a single gauge field was obtained in
Ref.[13], whose approach is different from ours.

\endpage

\chapter{ Low Energy Effective Lagrangian}

We now consider      self-dual systems with the specific potential (2.2).
Vortices at  rest are described by the configurations satisfying the
self-dual equation (2.7). As they are degenerate in energy,
there is no attractive or repulsive
forces  between them even though there may be velocity-dependent
 forces. We are interested in an
effective  action for slowly moving
vortices, which as a nonrelativistic action for interacting particles
consists of  terms  quadratic and linear in velocity.
 The linear term would represent the `dual magnetic' interaction between
vortices. The effective action also
should somehow take into account the fact
that vortices are not degenerate  in angular momentum.
If the kinetic energy of vortices is small enough, there would be very
small radiation and degrees of freedom except vortex positions may be
neglected. For this classical picture to be
consistent, quantum fluctuation
should be very small, which means $\kappa \gg 1$ and vortices interact
with each other strongly as argued in the previous section.

There are considerable works\refmark{10} for the effective action
for slowly moving vortices in
self-dual Maxwell Higgs systems. Their approaches are either
geometrical or numerical. Here we take somewhat different tactic
which seems to work also in the Maxwell Higgs case as shown in
appendix C, where we use the dual formulation even though the original
formulation would work equally well.
For self-dual Chern-Simons Higgs systems, the dual formulation seems
cumbersome for our present
purpose and we start from the original lagrangian
(2.1).

Consider  $n$ vortices with   uniform velocity
$\vec{u}$ and total mass $M = \pi v^2 n$. The field configuration for
this case can be obtained from that for vortices at rest by
a Lorentz transformation. We are interested in the
slow motion or nonrelativistic limit.  The $f$ field would transform
trivially and the gauge field  as vector would have a correction
linear in $\vec{u}$. The gauge fields  would satisfy the field
equation to  first  order in $\vec{u}$. We
can calculate the lagrangian
$L = \int d^2 r {\cal L}$ with this transformed configuration  and
 get the expected result $L = M\vec{u}^2/2 - M$.

For the field configurations   for the slowly moving
vortices of a given trajectory $\vec{q}_a(t)$,
 we imagine a generalization of the nonrelativistic
limit of the Lorentz transformation.
For the  consistency, we will assume that  there are first order
corrections to both scalar and gauge fields,  and require  that they satisfy
the field equations to  first order in the vortex velocities,
$\dot{\vec{q}}_a(t)$. The zeroth
order $f$ field would be $f(\vec{r};\vec{q}_a(t))$
 which satisfies Eq.(2.9).
The zeroth order $\theta$ field would be $\theta = \sum_a {\rm Arg}(
\vec{r}-\vec{q}_a(t))$. For a given  zeroth order scalar field, the
gauge field in the same order can be obtained from Eqs.(2.3) and (2.7).
$$ \eqalign{ &\ A_0(\vec{r};\vec{q}_a(t)) = -{1\over 2\kappa}
(f^2-v^2),\cr
&\ A_i(\vec{r};\vec{q}_a(t)) = -{1\over 2}\epsilon_{ij}\partial_j
\ln {f^2 \over \prod_a(\vec{r}-\vec{q}_a(t))^2 } . \cr}
\eqno\eq $$

{}From the lagrangian (2.1), we get the field equations,
$$ \eqalign{ &\ -\partial_\mu^2 f + (\partial_\mu \theta - A_\mu)^2 f
- U'(f) = 0 ,  \cr
  &\ \kappa F_{12} + f^2 (\dot{\theta} + A_0) = 0,\cr
&\ \kappa \epsilon_{ij} F_{0j} + f^2 (\partial_i \theta
+ A_i) = 0.\cr }
\eqno\eq $$
The field equation for the $\theta$ field can be obtained from
the  Jacobi identity for the gauge field equations.
We demand the field equations are satisfied to  first order in
the vortex velocities by the first order corrections, $\Delta f, \Delta
\theta, \Delta A_\mu$. We choose
 the gauge where $\Delta \theta =0$.  From Eq.(5.2), we get the
 first order field equations,
$$ \eqalign{ &\
\partial_i^2 \Delta f +     [A_0^2 - (\partial_i\theta + A_i)^2 ]
\Delta f +  2fA_0 (\dot{\theta}+ \Delta A_0) \cr
&\ \,\,\,\,\,\,\,\,\,\,\,
 - 2f (\partial_i \theta +A_i) \Delta A_i  - U''(f)\Delta f= 0, \cr
 &\ \kappa \epsilon_{ij} \partial_i
 \Delta A_j + f^2 (\dot{\theta} + \Delta A_0)
 + 2fA_0 \Delta f = 0,\cr
&\ \kappa \epsilon_{ij} (\dot{A}_j - \partial_j \Delta A_0) + f^2
\Delta A_i + 2f(\partial_i \theta + A_i)\Delta f = 0, \cr}
\eqno\eq $$
where  $f,A_\mu$ are given in zeroth
order.         We do not know at present moment
the solution $ \Delta f, \Delta A_\mu$ of Eq.(5.3)
 in terms of $f$ explicitly. If there is a unique solution of
Eq.(5.3) for a given trajectory of the vortices, we have definite
configurations  for slowly moving
vortices of  the self-dual Chern-Simons Higgs system.
For the  vortices of unform motion, the solution of Eq.(5.3) is
trivially given taking  the nonrelativistic limit of the Lorentz transformed
fields.

We imagine the field sum in the path
integral to be  restricted to these configurations for slowly moving
vortices. The field theoretic  action for slowly moving vortices then
becomes the effective  action  as a
functional of these vortex trajectories.
There will be terms linear and quadratic in the vortex velocities, but
no terms which just depend on the vortex positions.
We do not need to consider the  second order corrections  to the fields
to be consistent because their contribution
vanishes  due to the field equations satisfied by
the zeroth order fields.
By using Gauss's law (2.3), let us write the lagrangian density (2.1)
as
$$ \eqalign{
{\cal L} = &\ {1\over 2}\dot{f}^2 - \kappa \dot{\theta}F_{12} +
{\kappa \over 2} \epsilon_{ij} \dot{A}_i A_j \cr
&\ - {1\over 2}(\partial_i f)^2 - {1\over 2}f^2(\dot{\theta}+A_0)^2
- {1\over 2}f^2(\partial_i\theta  + A_i)^2 - U(f).\cr} \eqno\eq $$

The zeroth order term in the effective action can be calculated trivially
and becomes negative of the rest mass.   In Eq.(5.4) the last four terms
can be put into a sum of two squares plus the rest mass term as shown in
Eq.(2.5) and so yields only the second order terms. Thus  the first order
 term  $\Delta_1 {\cal L}$  is given by
$$ \Delta_1 {\cal L} = - \kappa \dot{\theta} F_{12} + {\kappa \over 2}
\epsilon_{ij} \dot{A}_i A_j,\eqno\eq $$
where the fields are given in  zeroth order.
With Eq.(5.2), the second part of the right-hand side of Eq.(5.5)
 is proportional to
$$ \eqalign{ 2\epsilon_{ij}\dot{A}_i A_j &\ =
- A_i \partial_i\partial_0 \ln {f^2 \over \prod}  \cr
&\ = -\partial_i \left\{  A_i \partial_0 \ln {f^2\over \prod}
    \right\} + (\partial_i A_i) \partial_0 \ln {f^2\over \prod}
   ,\cr} \eqno\eq $$
with the obvious understanding of $\prod$. Since the zeroth order gauge
efield is transverse and the boundary terms lead to no  contribution to
the effective action, Eq.(5.6) does not contribute to the effective
action. With $\dot{\theta} = \sum_a
\epsilon_{ij} \dot{q}^i_a \partial_j \ln |\vec{r}
-\vec{q}_a| $, the first part of the right-hand side of Eq.(5.5) is
proportional to
$$ \eqalign{\dot{\theta} F_{12}  = &\   \epsilon_{ij}\epsilon_{kl}
\dot{q}^i_a ( \partial_j \ln |\vec{r}-\vec{q}_a|) \partial_k A_l \cr
= &\ \dot{q}^i \partial_i( A_j \partial_j \ln |\vec{r}-\vec{q}_a| ) -
\dot{q}^i_a \partial_j( A_j \partial_i \ln |\vec{r}-\vec{q}_a|)
+ \dot{q}^i_a (\partial_j A_j) \partial_i\ln |\vec{r}-\vec{q}_a|
\cr
   &\  - \dot{q}^i_a \partial_j(A_i \partial_j \ln|\vec{r}-
\vec{q}_a|) + \dot{q}_a^i A_i \partial_j^2 \ln |\vec{r}-\vec{q}_a|
, \cr} \eqno\eq $$
where $\epsilon_{ij}\epsilon_{kl}=\delta_{ik}\delta_{jl} - \delta_{il}
\delta_{jk}$ is used and the sum over the indices $a$ is assumed.
 Now we can get the first order term of the effective lagrangian
from Eqs.(5.5) and (5.7),
$$\Delta_1 L = \int d^2 r \Delta_1 {\cal L} =
-2\pi \kappa \sum_a \dot{q}^i_a A_i (\vec{q}_a),\eqno\eq
$$
where $\partial_i^2 \ln |\vec{r}-\vec{q}_a| = 2\pi \delta (\vec{r}-
\vec{q}_a) $ is used and the boundary terms are dropped as they make no
contribution to the effective action.
There is another way to see that Eq.(5.8) is the only contribution
from Eq.(5.7) even though many terms in Eq.(5.7) seem singular at
vortex positions. Because $\dot{\theta} \sim 1/\delta $ and $F_{12} \sim
\delta^2$ with $\vec{\delta} = \vec{r}-\vec{q}_a$, $\dot{\theta}F_{12}$
vanishes at  vortex positions, which allows us to subtract these
points from the integration in Eq.(5.8). Then we can calculate the
integration with boundary contributions at vortex positions, getting the
 same result.

The second order term in the lagrangian would be
$$ \eqalign{ \Delta_2 {\cal L}   = &\ \,\,
   \kappa \Delta A_0 \Delta F_{12} + {1\over 2} \kappa \epsilon_{ij} (
   \Delta\dot{A}_i A_j + \dot{A}_i \Delta A_j ) \cr
   &\  + {1\over 2} \dot{f}^2
-{1\over 2}(\partial_i \Delta
f)^2 - {1\over 2}U''(f) (\Delta f)^2 \cr  &\ +
{1\over 2}f^2(\dot{\theta}+\Delta A_0)^2 - {1\over 2}f^2(\Delta A_i)^2
+{1\over 2}(\Delta f)^2 A_0^2 - {1\over 2}(\Delta f)^2 (\partial_i
\theta + A_i)^2 \cr
&\ +
2f\Delta f A_0 (\dot{\theta} + \Delta A_0) - 2f\Delta f (\partial_i
\theta + A_i )\Delta A_i .\cr
} \eqno\eq $$
First note that $\epsilon_{ij} \Delta \dot{A_i} A_j =
\epsilon_{ij} \dot{A}_i \Delta A_j$ up to a
total time derivative, which does not affect the effective action.
We use Eq.(5.2) to remove $U''$ and the Chern-Simons part up to total
derivatives. The resulting second order term is
$$ \eqalign{ \Delta_2 {\cal L} = &\
{1\over 2} \dot{f}^2 + {1\over 2} f^2
(\dot{\theta}+ \Delta A_o)^2 + {1\over 2} f^2 (\Delta A_i)^2   \cr
&\ +
f  \Delta f A_0 (\dot{\theta} + \Delta A_0) + f \Delta f
(\partial_i \theta + A_i)  \Delta A_i \cr}
\eqno\eq $$

Hence, we have obtained the effective action  for slowly moving
vortices. From Eqs.(5.8) and (5.10), we can see that
the effective lagrangian for slowly moving vortices is given by
$$  \eqalign{
L_{{\it eff}} (\vec{q}_a, \dot{\vec{q}}_a) =
\int d^2 r &\ \,\,\, \biggl\{
{1\over 2} \dot{f}^2 + {1\over 2}f^2 [( \dot{\theta}+
\Delta A_0)^2 +  (\Delta A_i)^2]  \biggr. \cr
   \,\,\, &\ \biggl.    +
 f\Delta f A_0 (\dot{\theta} + \Delta A_0) + f \Delta f
 (\partial_i \theta + A_i) \Delta A_i  \biggr\} \cr
     - 2\pi \kappa &\ \sum_a
 \dot{q}_a^i A_i(\vec{q}_a). \cr}  \eqno\eq $$
Here $f, A_i$ are given by Eqs.(2.9) and (5.2)
 and functions of $\vec{q}_a$'s. $\Delta f$ and
$\Delta A_\mu$ are given by the first order field equation (5.3)  and
is linear in velocity.
  The effective lagrangian is made of the usual quadratic
terms and the linear terms which  describes the magnetic interactions
between vortices.

We made some reasonable assumptions to derive the effective action
for slowly moving vortices. A configuration for moving vortices
is specified by $f+ \Delta f, A_\mu + \Delta A_\mu$.  The
energy functional for this configuration consists of the rest mass and
terms linear and quadratic in the  vortex velocities. From Eq.(5.3),
One can easily show that $\Delta_1 E = \int d^2 r \kappa \epsilon_{ij}
\partial_j [ A_0 \Delta A_i]$, which vanishes. The
 quadratic terms in the energy
functional is not identical to the quadratic part of the effective
lagrangian. The difference is
$$ \eqalign{  \Delta_2 E - \Delta_2 L  = \int d^2  r &\
\biggl\{ f\Delta f A_0 (\dot{\theta} +
\Delta A_0) + f\Delta f (\partial_i \theta + A_i) \Delta A_i \biggr.
  \cr  &\ \biggl.
+ {1\over 2} (\partial_i \Delta f)^2  + {1\over 2} (\Delta f )^2 A_0^2
+ {1\over 2}(\Delta f)^2 (\partial_i \theta + A_i)^2 \biggr\},
\cr}
\eqno\eq $$
which does not seem to  vanish.  We believe that the quadratic part of
the effective action is given by $\Delta_2 L$ rather than $\Delta_2 E$
because the linear part cannot be obtain from the energy point of view.
For uniformly moving vortices, the first order correction (5.3) of the
fields  would
be given by the nonrelativistic limit of the Lorentz transformationed
fields and the effective lagrangian becomes the total kinetic energy of the
system.

We can use Eq.(3.13) to express the linear term in terms of the
dual gauge field. As there is no external charge, we can choose
the gauge where $V_\mu = 0$ and $A_\mu = -H_\mu/2\pi \kappa$.
The linear term (5.8) becomes then
$$  \Delta_1 L  =    \sum_a \dot{q}^i_a H_i(\vec{q}_a),  \eqno\eq $$
which is exactly what we get from the dual formulation and need
for the statistics of vortices. The linear part (5.13) implies the
`dual magnetic' interaction between vortices. In Sec.~4, we argued
that dual magnetic interaction originates both ordinary magnetic and
Magnus forces. This linear interacting terms (5.13) leads to the
statistical phase between vortices.
In Ref.[14], Eqs.(5.5) and (5.6) have been examined to get
the   statistics of vortices at large separation but was not put into a
simple form as Eq.(5.8) or (5.13), let alone its physical meaning.

{}From Eq.(2.3) we can see the
 original gauge field strength $F_{12}$ vanishes at $\vec{r}=\vec{q}_a$.
This does not mean that the field strength felt by vortices vanish.
The reason is that the gauge field $A_\mu(\vec{r}; \vec{q}_b)$
as a function of $\vec{q}_a$ when $\vec{r}=\vec{q}_a$
is different from that as a function of
$\vec{r}$. From Eqs.(2.10) and (5.1), we can  get the field strength felt
by the vortex at $\vec{q}_a$,
$$ \eqalign{ {\bf H}_{12}(\vec{q}_a)
 &\ = -2\pi \kappa \epsilon_{ij}
{\partial A_j \over \partial q_a^i}  \cr
&\ = - 2\pi \kappa  {\partial b^i_1 [\vec{q}_a] \over
\partial q^i_a }. \cr} \eqno\eq $$
Similar consideration would apply as well to the cases studied in
appendices A and B.

There is an interesting check of the linear term. Let us consider
the total angular momentum of vortices from the low energy effective
lagrangian (5.11),
$$ \eqalign{ J_{orbit} &\ = \sum_{a} \epsilon_{ij} q_a^i
{\partial  L\over \partial \dot{q}_a^j} \cr
&\ = - 2\pi \kappa
\sum_{a} \epsilon_{ij} q^i_a A_j(\vec{q}_a) + O(\dot{q}_a^i).\cr}
\eqno\eq $$
With Eqs.(2.10) and (5.1), one can see that
$$ J_{orbit}
=  -2\pi\kappa \sum_a  \vec{q}_a\cdot \vec{b}_1[\vec{q}_a]
+ O(\dot{q}^i_a),\eqno\eq $$
which is identical to the orbital part in Eq.(2.19) for vortices at rest.
Our effective lagrangian for slowly moving vortices is consistent with
the field theory lagrangian.

Let us briefly study  the dynamics of slowly moving two votices.
First consider two overlapped vortices with a small initial kinetic energy.
The initial angular momentum would be very close to
$4s_v$. If they can escape
from each other to the
spatial infinity, their angular momentum would be
the sum of spins and orbital angular momentum, $2s_v + u_0 b$,
where $u_0 $ is the asymptotic speed and $b$ is the impact parameter.
As we can choose the kinetic energy, or $u_0$,
arbitrarily small, the angular
momentum conservation says that the impact parameter becomes
arbitrary large, which is impossible because the force is short
ranged. Rather, we think that two vortices are bound together by
the mutual magnetic field. By turning around this argument, one can also see
that two vortices from the
spatial infinity with very small kinetic energy
cannot make a head-on collision rather they will always veer off from each
other.

In the center of mass frame of two vortices, their positions are given
by $\vec{q}_1 = -\vec{q}/2, \vec{q}_2 =\vec{q}/2$. The scalar field
configuration would be symmetric under the inversion, that is,
$f(\vec{r};\vec{q}_a) =  f(-\vec{r};\vec{q}_a)$, which with Eq.(2.10)
means  that $\vec{b}_1[\vec{q}_1 ] = - \vec{b}_2[\vec{q}_2]$.
There is a rotational symmetry of the effective action under
the rotation of $\vec{q}$, which implies that
$$ \vec{b}_1 =
\hat{q} B_1(q) + \hat{z}\times \hat{q} B_2(q), \eqno\eq $$
where $q = |\vec{q}|$ and $(\hat{z}\times \hat{q})^i = \epsilon_{ij}
q^j$.  Eqs.(2.19) and (5.17) imply  that the total angular
momentum for two vortices at rest is  $J = -2\pi\kappa qB_1(q) -
2\pi\kappa$. The linear action (5.13) for two vortices in the center
of mass frame becomes $\Delta_1 L = - 2\pi \kappa \dot{q}^i
A_i(\vec{q}_1)$. The magnetic field felt by the reduced one body would
be then
$$ \eqalign{ {\bf H}_{12}(q)  &\ = -{2\pi\kappa \over q}
{ d(qB_1) \over dq} \cr
&\ =   {1\over q}{ d J(q) \over  d q}.  \cr }
 \eqno\eq $$
The magnitude of the angular momentum decreases with the separation
between vortices. The sign of the magnetic field is then opposite to
the sign of the angular momentum, which in turn leads the direction
to which the vortex trajectories bends. When the vortex spin is
positive, or, $\kappa <0$, vortices in a two vortex system
turn right. For the negative spin, vortices turn left.
 From this one can easily obtain a
qualitative pictures of the dynamics of two vortices which are
either bounded closely or starting and ending at the spatial infinity.
However, the detail pictures seem to be complicated  and will not be
pursued here.

Somewhat similar behavior has been observed numerically
in another kind of
self-dual system  with global charge and topology in three
dimensions.\refmark{15}
Our approach may shed some light on the physical understanding of the
interaction between those solitons.

Finally, let us consider the meaning of the effective lagrangian (5.11).
We do not have any geometric derivation of the quadratic term,  but we
can take the quadratic term as a metric on the moduli
space, the space of the self-dual configurations of vortices.
 The linear term
could be interpreted as a magnetic field in the moduli space.
Vortices are then moving along
     geodesics  determined  by the metric and
magnetic field.    Vortices carry spin and   may feel the spin
connection of the metric on the moduli space.
 Since the spin connection
could be interpreted as a sort of the gauge field, the linear term in our
effective lagrangian may be interpretable as the spin connection,
making the effective action fully geometric.
 To see
this, we need a better understanding of the quadratic part of the
effective action.

\endpage

\chapter{Conclusion}

We understand now various aspects of vortex dynamics in
Chern-Simons Higgs systems. We have the dual formulation in the path
integral formalism, where the interaction between vortices manifests.
  The statistics of vortices
comes from the Aharanov-Bohm phase of the dual gauge interaction, which
combines the usual electromagnetic and Magnus forces.
In the dual formulation, we included
the external field and current, which could be dynamical. In
self-dual models we studied the properties of static  vortices and
presented an effective action for slowly moving vortices.

There seems to be some interesting directions to take from here.
 One direction is to find the
 further use of the dual formulation. We can ask whether the perturbative
expansion is possible in the dual formulation. For vortices
 moving on a curved surface whose typical length scale is much
larger than the size of vortices, there could be an force on vortices
via spin connection because vortices carry spin. Maybe our approaches
would shed some light on that.  It would be also interesting to find whether
there is a dual formulation of the nonrelativistic limit of the theory
in the symmetric phase.
 Another is to understand better
 the effective action for slowly moving vortices
 and its dynamical consequences.
 Besides the statistics, we have not studied  the quantum aspects
of vortex dynamics. Quantum aspects of vortices in the field theoretic
and effective action levels need further investigation.

\vskip 0.6in
\centerline{\bf Acknowledgment}
This work is supported in part by the Korea Science and Engineering
Foundation (Y.K.), the NSF under Grant No. PHY89-04035,
Dept. of Energy (Y.K and K.L.), the NSF
Presidential Young Investigator Award (K.L) and an Alfred P. Sloan
Fellowship (K.L.).  K.L. would like to thank the organizers of the
Cosmic Phase Transition Workshop at ITP, U.C. Santa Barbara,  the
Center for Theoretical Physics of Seoul National University, and Aspen
Center for Physics where a  part  of this work was done.

\endpage

\Appendix{A.}

Here we derive  the Magnus force in the simplest example.  Consider
the theory of a complex scalar field in three dimensions with a global
$U(1)$ symmetry. The  lagrangian is given by
$$ {\cal L} = |\partial_\mu \phi|^2 - U(\phi).\eqno\eq $$
The generating functional is
$$ Z = <F|e^{-iTH} |I> = \int [d\phi] [d\phi^*]
 e^{i \int d^3 x {\cal L}}.\eqno\eq $$
With $\phi = f e^{i\theta}/\sqrt{2}$,
the lagrangian becomes
$$ {\cal L} = {1\over 2} (\partial_\mu f)^2 + {1\over 2}f^2 (\partial_\mu
\theta)^2 - U(f).\eqno\eq $$

The conserved current for the global abelian symmetry is $j_\mu = f^2
\partial_\mu \theta$. Suppose we are interested in the minimum energy
density configuration for a given uniform charge density $j^0 =
\rho_B$. The phase becomes $\theta = wt$ with a constant $w$ and the
$f$ field is fixed by the minimizing the energy density,
$$ U_{\it eff}(f) = {1\over 2f^2} \rho_B^2 + U(f).\eqno\eq $$

There could be a global vortex with this background.
The ansatz will be $f(r)$ and $\theta = wt + n\varphi$. These vortices
carry logarithmically divergent energy and quadratically divergent
angular momentum when the charge density is non-zero.

Let us consider the motion of vortices and antivortices with some background
charge. For example, one is imagining some bosonic superfluid or $Q$-matter.
In the same way as in Sec.~3 we introduce an auxiliary field $C^\mu$
to linearize the second term of the lagrangian (A.3). Separate the
phase $\theta$  into a part for vortex configurations and a
part for
single-valued fluctuations as in Eq.(3.3). Integrate
over the fluctuation to get a new gauge field $H_\mu$ for $C^\mu$.
After some further steps similar as in
      Sec.3, we arrive at the dual
formulation of the generating functional,
$$ Z = \int [df][dH_\mu][dq^\mu_a]  e^{i\int d^3 x {\cal L}_D },\eqno\eq $$
where
$$ {\cal L}_D = {1\over 2} (\partial_\mu f)^2  - {1\over 16\pi^2 f^2}
 H_{\mu\nu}^2  + H_\mu K^\mu.\eqno\eq $$
with $K^\mu$ given in Eq.(3.5).
There is no Jacobian factor in measure as in Sec. 3.
In the dual formulation the Goldstone boson is described by the massless vector
field with the Maxwell kinetic term. Vortices become charged particles
and the logarithmically divergent self-energy comes from the divergent
Coulomb energy.

In the dual formalism, the  conserved current for the global symmetry becomes
$$ j^\mu = f^2 \partial^\mu \theta = \epsilon^{\mu\nu\rho}\partial_\nu
H_\rho.\eqno\eq $$
The uniform charge density background becomes a uniform magnetic field
background. Vortices  moving on a  uniform charge background  would feel
the Magnus force as a Lorentz force in the dual formulation.

Let us now do a little bit of fluid dynamic approach to the
Magnus force to figure out the direction.
For the positive $w $ and
$ n$, the momentum density  flow $T^{0i} = - \int d^2r [
\dot{f}\partial_i f + f^2 \dot{\theta}\partial_i \theta ]$
around the vortex is clockwise,
resulting in the negative angular momentum density. Let us consider
a vortex moving to the negative $x$ axis. This is very similar to the
case where the 2-dimensional baseball moving in the same direction with
the same rotation, feeling the net Magnus force  in the positive $y$
direction. This direction of force is exactly that of the Lorentz
force  one would get from the  dual lagrangian (A.6)

\endpage

\Appendix{B.}

Here we present the path integral derivation of the
dual transformation for a    Maxwell Higgs theory in three
dimensions.
The lagrangian for a complex scalar field $\phi=fe^{i\theta}/\sqrt{2} $
coupled to the gauge field $A_\mu$ is
$$ {\cal L} = -{1\over 4 e^2 }F_{\mu\nu}^2 +
{1\over 2} (\partial_\mu f)^2 + {1\over 2}f^2 (\partial_\mu
\theta + A_\mu + A^{ext}_\mu )^2 - U(f)  + A_\mu J^\mu,\eqno\eq $$
where $J^\mu$ is the external current and $A^{ext}_\mu$ is the
external gauge field.
As in  Sec.3,  we introduce
the vortex current  $K^\mu $ and integrate over the fluctuation part
of the  $\theta$ field,
resulting in a dual gauge field $H_\mu$. The effective lagrangian becomes
$$ \eqalign{  {\cal L}' = &\ {1\over 2}(\partial_\mu f)^2 -U(f)
 - {1\over 16\pi^2 f^2}H_{\mu\nu}^2 + H_\mu K^\mu
-{1\over 4 e^2 }F_{\mu\nu}^2 \cr
&\  + {1\over 4\pi}\epsilon^{\mu\nu\rho}H_\mu F_{\nu\rho}
 + {1\over 4\pi}\epsilon^{\mu\nu\rho}H_\mu F^{ext}_{\mu\nu} + A_\mu
J^\mu . \cr}  \eqno\eq $$

In order to treat $F_{\mu\nu} $ and $A_\mu$ to be independent from each other,
we introduce an vector  field $N_\mu$ so that
$$ \eqalign{ \int [d &\ F_{\mu\nu}][  dA_\mu ]  \delta(F_{\mu\nu} -
 (\partial_\mu A_\nu -
\partial_\nu A_\mu))... \cr
&\ = \int [dF_{\mu\nu}][dA_\mu][dN_\mu]
\exp \{ i \int d^3x {1\over 4\pi}
\epsilon^{\mu\nu\rho} N_\mu [ F_{\nu\rho} - (
\partial_\nu A_\rho - \partial_\rho A_\nu) ] \}... \cr}  \eqno\eq $$
The $F_{\mu\nu}$ integration is just a gaussian integral and so trivial.
The $A_\mu$ integration leads to a factor
$$ \delta( {1\over 2\pi}\epsilon^{\mu\nu\rho} \partial_\nu N_\rho -
J^\mu ),\eqno\eq$$
which is consistent only if the external current $J^\mu$ is conserved
explicitly. Thus, unlike to Sec. 3 the external current could have a
dynamical origin only for the case when  $A_\mu J^\mu$  is  replaced by
$$ {\cal L}_W = \epsilon^{\mu\nu\rho} A_\mu \partial_\nu W_\rho + {\rm
kinetic\, terms}.\eqno\eq $$
The external gauge field can be made dynamical by simply replacing
$A^{ext}_\mu$ by, for example, $W_\mu$ in Eq.(B.5)

Let us consider a single-valued  scalar field $\zeta$ such that
$$ N_\rho =  \bar{N}_\mu + \partial_\rho \zeta,\eqno\eq $$
where
$$ \epsilon^{\mu\nu\rho} \partial_\nu \bar{N}_\rho  = 2\pi J^\mu \eqno\eq $$
and $\partial_\rho \bar{N}_\rho = 0 $. A  uniform external electric
charge density corresponds to a uniform magnetic field in the vector potential
$\bar{N}_\rho$. If we put $\bar{N}_\rho =\partial_\rho \bar{\zeta}$ for
a point current of unit charge,    $\bar{\zeta}$ becomes
multivalued with shift $2\pi$, which can be absorbed into $\zeta$. This
allows an interpretation that point external charges of integer
charge  are    vortices in the $\zeta$ variable.

Putting together, the generating functional after the dual transformation
becomes
$$ Z = \int [df][dH_\mu][dq^\mu_a][ d\zeta] \delta(\epsilon^{\mu\nu\rho}
\partial_\nu \bar{N}_\rho  - 2\pi J^\mu) \exp \{ \int d^3x {\cal L}_D \},
\eqno\eq $$
where
$$ \eqalign{  {\cal L}_D = &\  {1\over 2}(\partial_\mu f)^2 -U(f)
   -{1\over 16\pi^2 f^2} H_{\mu\nu}^2 + H_\mu K^\mu \cr
&\ + {e^2\over 8\pi^2} (H_\mu + \bar{N}_\mu + \partial_\mu \zeta)^2 +
{1\over 4\pi}\epsilon^{\mu\nu\rho}H_\mu F_{\nu\rho}^{ext}.\cr}\eqno\eq $$
with $K^\mu$ given in Eq.(3.5).
There is an obvious abelian gauge symmetry in the dual lagrangian. The
point external currents of integer charge could appear as vortices in
the $\zeta$ variable.  The massive vector bosons of spin $\pm 1$ are
described by the Maxwell Higgs terms in both formalisms.

If there is a uniform electric charge density background, we know that
there should be a uniform charge density background of the opposite
charge carried by the Higgs field to have a finite Coulomb energy. In
the dual formulation, there is a uniform external magnetic field
carried by $\bar{N}_\rho$ which should be balanced by the unform
magnetic field of the opposite sign carried by $H_\mu$ for a finite
energy density as one can see in the dual lagrangian (B.9).
In the dual formulation vortices moving on the uniform charged
background are  equivalent to  charged particles moving on a uniform
external magnetic field and vortices feel the Magnus force as an
effective Lorentz force.

\endpage

\Appendix{ C.}

Here we study the effective lagrangian for slowly moving vortices  in
self-dual Maxwell Higgs systems  in the dual formulation of appendix B.
The self-dual model is fixed by choosing  the  potential,
$$ U(f) = {e^2\over 8} (f^2-v^2)^2.\eqno\eq  $$
The energy functional of the dual lagrangian (B.9) can be rewritten as
$$ \eqalign{ E = \int d^2 r &\  \left\{ {1\over 2}\dot{f}^2 +
{1\over 8\pi^2 f^2} H_{12}^2 +
 {e^2 \over 8\pi^2} (H_i + \bar{N}_i +
 \partial_i \zeta)^2  \right.
\cr
&\ \left. + {1\over 2} (\partial_i f \pm {1\over 2\pi f } H_{0i})^2 +
{e^2 \over 8\pi^2} (H_0+\bar{N}_0+ \partial_0 \zeta\mp \pi(f^2-v^2))^2
\right\} \cr
 \pm \pi v^2 n &\ ,   \cr } \eqno\eq $$
where the vorticity $n= \int d^2 r K^0 $ appears because of Gauss's
 law,
$$ \partial_i ( {1\over  f^2}H_{0i} )  + e^2 (H_0+
\partial_0 \zeta) + 4\pi^2 K^0 = 0.\eqno\eq $$
The energy is bounded, $E \ge \pi v^2 |n| $.  As there is no external
charge and field, we choose the gauge where  $\bar{N}_\mu =0 $ and $\zeta=0$.
The energy bound is saturated by the configurations satisfying
                 $\dot{f}=0, H_i = 0$,
$$ \eqalign{ &\  H_0  = \pm \pi (f^2 -v^2),\cr
&\ H_{0i} = \mp \pi \partial_i f^2, \cr} \eqno\eq $$
and Gauss's law (C.3).  Two equations in Eq.(C.4) are consistent to
each other. Eqs.(C.3) and (C.4)  can be put together into an  equation for $f$,
$$ \partial_i^2 \ln f^2 - e^2 (f^2 -v^2) = 4\pi \sum_a \delta(\vec{r} -
\vec{q}_a).\eqno\eq $$

Let us try to derive the low energy effective lagrangian in the dual
formalism.  We know how the fields transform under the nonrelativistic
limit of the Lorentz transformation of all vortices. The scalar field
will be invariant but there is a nontrivial correction $\Delta H_\mu$
to the gauge field. The gauge field satisfies the Maxwell equation in
the first order of vortex velocity  $\dot{\vec{q}}_a$,
$$ \partial_\nu ({1\over f^2}H^{\nu\mu}) + e^2 H^\mu + 4\pi^2 K^\mu =
0.\eqno\eq
$$

For  slow moving vortices with vortex positions
$\vec{q}_a(t)$,  we  assume that the fields
transform like a complicated version of the Lorentz transformation.
The scalar field would be given simply as $f(\vec{r};\vec{q}_a(t))$.
There would be a correction to the gauge field linear in the velocity.
We require that the Maxwell equation is again satisfied to first
order in velocity. Note that the velocity of vortices would appear
explicitly in the Maxwell equation by the current $K^i = \sum_a
\dot{q}_a^i \delta(\vec{r}-\vec{q}_a)$.

In  zeroth order,  only $H_0$ is nonzeros as one see from
Eq.(C.4). The first order part of Eq.(C.6)  is
$$ \eqalign{ &\ \partial_i({1\over f^2}\partial_i
\Delta H_0) + e^2 \Delta H_0 =0,\cr
&\  -\partial_0({1\over f^2}\partial_i  H_{0}) +
\epsilon_{ij}\partial_j({1\over f^2} \Delta H_{12}) + e^2 \Delta H_i =
4\pi^2 K^i . \cr} \eqno\eq $$
The first part of Eq.(C.7)  implies that $\Delta H_0=0$.
For the second part  of Eq.(C.7), we apply both
 $\partial_i $ and $\epsilon_{li}\partial_l$, leading to
$$ \eqalign{ &\ \partial_i \Delta H_i = \pi \partial_0 f^2,\cr
&\ \partial_i^2({1\over f^2} \Delta H_{12}) - e^2 \Delta H_{12} = 4\pi^2
\sum_a  \epsilon_{ij} \dot{q}_a^i \partial_j \delta(\vec{r}-\vec{q}_a).
 \cr}
\eqno\eq $$
As we know the divergence and curl of $\Delta H_i$, in principle we
can find $\Delta H_i$ explicitly.

Before we consider the effective action, let us ask whether the $f$
field satisfies  its field equation to  first order in the vortex
velocity. One can be easily convinced that the first order correction
$\Delta f$ can be put to be zero consistently.

We now the field configuration of slowly moving vortices for a given
trajectory. Let us calculate the field theory action  from Eq.(B.9)
for these
configurations. The zeroth order term is  minus of the rest mass.
There is no first order term.  The second becomes the  effective  action
for slowly moving vortices. The field equation (C.8) is essential in this
derivation. The effective lagrangian is
$$ L_{{\rm eff}} (\vec{q}_a, \dot{\vec{q}}_a) =
\int d^2 r \left\{ {1\over 2}\dot{f}^2 + { 1\over 8\pi^2 f^2}(\Delta
H_{12})^2 + {e^2 \over 8\pi^2}(\Delta H_i)^2 \right\}.\eqno\eq $$
Let us see what happens in the original formulation.
The field equations in the intermediate lagrangians imply
$$ \eqalign{ &\ \epsilon^{\mu\nu\rho}\partial_\nu H_\rho
= 2\pi f^2 (\partial^\mu \theta +  A^\mu),\cr
&\ \epsilon_{\mu\nu\rho} H^\rho = 2\pi e^2 F_{\mu\nu},\cr} \eqno\eq $$
which implies that
$$ \eqalign{ {e^2 \over 2\pi }\Delta H_i &\ =
 \epsilon_{ij}\dot{A}_j \cr &\
 = \partial_i\partial_0 \ln f
- \epsilon_{ij}\partial_j \theta  \cr} \eqno\eq $$
in the $A_0=0$ gauge. With this identification, our effective lagrangian
(C.9) can be shown easily to be identical
to that of Samols' in Ref.[10].

\endpage

\refout
\endpage
\figout
\end